# Physics at the University of Lviv: the first two centuries in the bibliographic aspect

*Andrij Rovenchak*

Department for Theoretical Physics, Ivan Franko National University of Lviv
12, Drahomanov St., Lviv, UA-79005, Ukraine
tel: +380 (32) 2614443
e-mail: andrij.rovenchak@gmail.com

**Abstract**. A detailed bibliography related to physics at the University of Lviv (Leopolis, Lemberg, Lwów) in 18th–19th centuries is presented. Over ninety works of various types are listed with a large share being illustrated by title or starting pages. Brief biographical accounts of the authors are given to put their works in the context of the University history.

# Physics at the University of Lviv: the first two centuries in the bibliographic aspect

*Andrij Rovenchak*

Department for Theoretical Physics, Ivan Franko National University of Lviv, Ukraine

**Abstract**. A detailed bibliography related to physics at the University of Lviv (Leopolis, Lemberg, Lwów) in 18[th]–19[th] centuries is presented. Over ninety works of various types are listed with a large share being illustrated by title or starting pages. Brief biographical accounts of the authors are given to put their works in the context of the University history.

## 1. Introduction

Lviv (Leopolis in Latin, Lwów in Polish, Lemberg in German) is a city in the Western part of Ukraine. It hosts the University, which was established in 1661 being thus one of the oldest continuously operating academic institutions in Eastern Europe. Founded as a Jesuit Academy by a decree of John II Casimir (Jan II Kazimierz Waza), a Polish King, the University was later an Austrian, Polish, Soviet, and finally Ukrainian institution of higher learning reflecting a turbulent history of this region in the 17[th]–20[th] centuries [cf. Woleński 1997].

Various organizational and administrative changes, however, did not affect the teaching of basic disciplines significantly. Physics, as a part of Aristotle's philosophy, was taught at the University of Lviv, most probably, since the very beginning. For instance, a manuscript entitled *Philosophia naturalis seu Physica disputationibus illustrata anno 1688 die 17 Februarii inchoata Leopoli sub reverendo patre Georgio Gengell* is mentioned in [Matwiiszin 1974], syllabuses of the philosophy courses by Jan Wojnarowicz in 1714/15–1715/16, Tomasz Dunin in 1718/1719–1719/1720, and by an unknown professor in 1724/1725–1725/1726 contained explicitly mentioned parts on physics [Symčyč 2009: 233–237]. Interestingly enough, initially the teaching of physics was considered illegal for the Jesuit Academy in Lviv (as opposed to the Cracow Academy) and was banned at least twice, in 1708 and 1728, by decrees of August II the Strong, a Polish King [Finkel & Starzyński 1894: I 25].

In the paper, the bibliographic information related to physics at the University of Lviv is collected. The time span covers the period till 1872 when the Chair of Physics was split into two separate chairs, Experimental and Theoretical Physics, respectively. For the latter, several bio-bibliographic studies have been published already [Rovenchak 2009; 2012; 2013]. An attempt is made to preserve the data on the author, title, place of publication, etc. as exact as possible. For instance, instances of "Dr." or "Professor" are kept together with author's name as they appear in the original; abbreviations and omissions are avoided in longer titles, especially in Latin.

Short biographic accounts of authors are presented if known. After 1783, when the Chair of Physics was created, all the known publications of persons being the professors (and in later years also docents) at this chair are listed, even if these belong to other disciplines or are not scientific by nature at all. In some cases, it is not possible to define the affiliation of authors and the attribution to the University of Lviv is based on secondary sources.

## 2. Before the Chair of Physics

By the middle of the 18[th] century, the development of natural sciences made clear the future shaping of physics as a separate discipline. During this period, first publications with elements of physics appeared at the Lviv Jesuit Academy [Bednarski 1936; Bieńkowska 1971:132; Sokolov 2000].

As the study of bibliographic databases and indices [Kotula 1928; Zapasko & Isajevyč 1981–84] demonstrates, the first title bearing the sign of new trends was probably the following collections of theses:

1. * **Scientia naturalis in suis thesibus a Leopoliensi Soc. Jesu**. — Leopoli : S. J., 1744. ESTREICHER 27, 310.
(Here and further, an asterisk after the number means that the respective item was not checked *de visu*. For such titles, the references to some standard bibliographies or library catalogs are given in small caps. All the other publications were either verified at the Lviv University Library (LUL) and Lviv Stefanyk Scientific Library (LSL) and/or downloaded from the digital collections listed in the References at the end of the paper.)

The year of the above publication coincides – and this is not probably entirely by chance – with the year the Chair of Mathematics was established at the University of Lviv.

Another title from this period is

2. **Universa Mathesis Terræ Cæliq; Considerationi Duorum Annorum Cursu Impensa** : Ad utilitatem Reipublicæ Polonæ seu pace, seu bello promovendam, ad rerum familiarium, atq; ad vitæ humanæ commoda ordinata demum in Scholis Palatinis Collegii Leopoliensis Societatis Jesu Publicis demonstrationibus exposita Majoriq; Dei gloriæ nec non B.M.V.S.P.N.I.X.B.A.K.SS.OO. Cultui dicata Anno Domini CIƆIƆCCXLV. — [Leopoli : S. J., 1745]. — [16] p.

This was an announcement on public demonstrations to be held at "Aula Scholarum Collegii Leopoliensis S.I." in June, 1745. Topics of the demonstrations included in particular: *Ex machica, Ex phonurgia, Ex optica, Ex catoptrica, Ex dioptrica. Ex astronomia sphærica, Ex astronomia theorica,* etc.

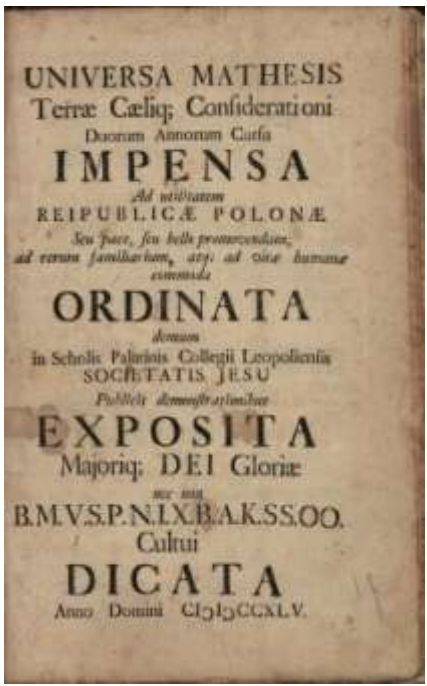
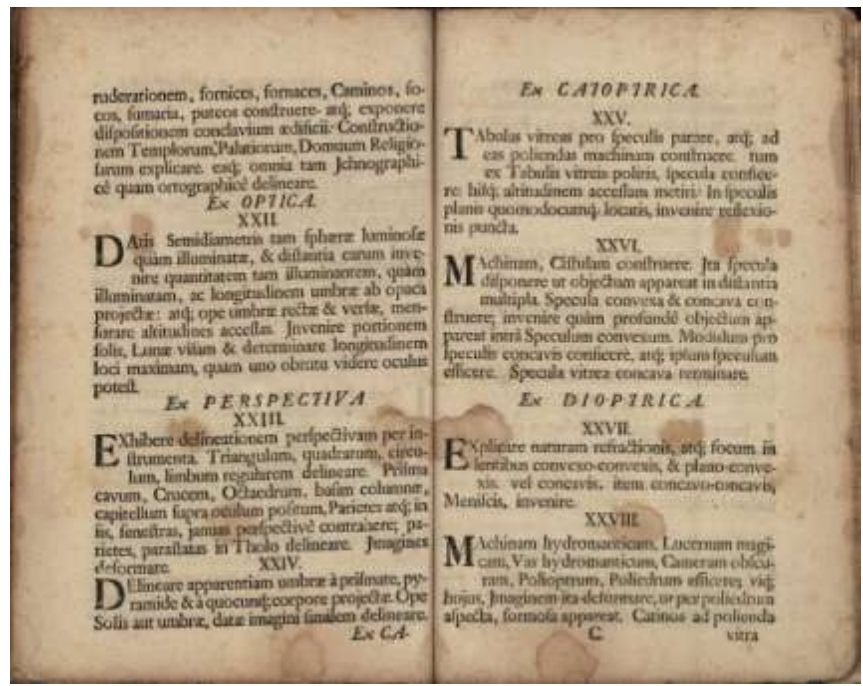

Figure 1: Title page and excerpt from *Universa Mathesis…* (item 2) [Image source: Polona]

The Lviv University library holds an anonymous manuscript entitled *Philosophia peripatetica Tribus partibus Logica Physica Methaphysica Comprehensa veterum principijs Recentiorum Experimentis Conformata [Peripatetic philosophy in three parts of logic, physics, and metaphysics on the basis of early principles and moderns experiments]*, see Fig. 2. The manuscript, which could be a collection of lecture notes, is dated 1747.

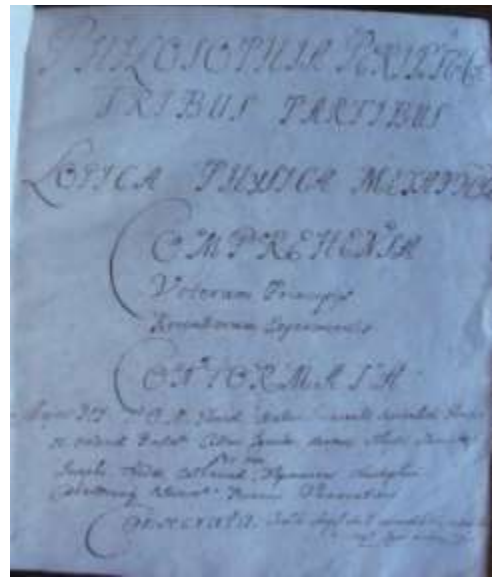

Figure 2: Title page of the manuscript *Philosophia peripatetica…* [Image source: LUL]

Since physics was a part of the philosophy course at the University, some works by students and graduates with elements of physics were published, in particular:

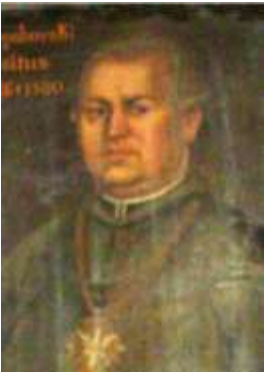

3. **Propositiones Ex universa philosophia** / à Perillustri & Admodum Reverendo Domino *Cajetano Tagoborski* Vexilliferi Czerniechoviensis Filio In Scholis Premisliensibus S. Theologiæ Auditore dum In Universitate Leopolitana Magisterio Philosophiæ donaretur propugnatæ Anno Domini 1761. Mense Junio.— [Leopoli: S. J., 1761].— [26] f., 4 leaves of illustrations.

**Kajetan (Cajetanus Bonawentura) TĘGOBORSKI** (\*ca. 1720 – †after 1800?) in the years 1776–1800 was a scholastic and provost of the cathedral chapter in Przemyśl, diocese administrator *sede vacante* [ESTREICHER 31, 76]. Photo by Arkadiusz Bednarczyk (from the portrait in the sacristy of the Archicathedral in Przemyśl).

4. **Synopsis Philosophiæ recentioris in Universitate Leopoliensi traditæ**: Ipso biennio Scholastici decursu / ad disserendum publice proposita A. P. *Ludovico Hoszowski* Soc. Jesu in eadem Universitate Publico Philosophiae Professore.— Leopoli: Typis Academicis, 1766.— [32] f., 1 leaf of illustrations.

**LUDWIK HOSZOWSKI** (\*1732 – †1802) was a Professor of Mathematics at the Jesuit Academy in Lviv from 1769 to 1773. In 1771–73, he was also a Professor of Astronomy and Prefect of the Mathematical Museum in the Academy. After the suppression of the Jesuit Order, he left for an ecclesiastical post in Przemyśl [Domoradzki 2011: 82; Aspaas 2012: 165 with reference to Fischer 1984].

Two more publications related to physics are known from 1773, namely:

5. \* **Dissertationes ex optica aerometria mechanica et astronomia ipso Studij Biennalis decursu ab auditoribus Matheseos publice habita in Collegio Leopolitano P. P. S. J. anno 1773. Mense April**.
   — Leopoli : S. J., 1773.
   ESTREICHER 15, 235.

6. \* **Propositiones ex universa philosophia publicae disputationi expositae in Universitate Leopoliensi Soc. Iesu anno 1773**. — Leopoli : S. J., 1773.
   ESTREICHER 25, 289.

Upon the First Partition of Poland in August, 1772, Lviv became a part of the Austro-Hungarian Empire. Soon after that, Pope Clement XIV issued two briefs (Lat. *breve*): in the first one, *Dominus ac Redemptor noster* (on 21 July 1773), the suppression of the Society of Jesus was declared; the second brief, *Gravissimis ex causis* (on 16 August 1773) dealt with practical issues of the suppression [Finkel & Starzyński 1894: I 30]. The Lviv Jesuit Academy ceased to exist in September, 1773.

# 3. Josephian University

While the order of Jesuits was dissolved and all its structures, including educational units, were closed, the former Jesuit Academy in Lviv was converted into the State Academy called *Theresianum*. At that time the reformation of education in the Habsburg Empire was under way. In the Academy, the teaching of disciplines remained roughly on the previous level. The preparation to a new educational institution started almost immediately.

From that time, several publications are known.

7. **Ephemerides astronomicæ Anni Bissexti 1776 ad Meridianum Vindobonensem**: cum Appendice observationum astronomicarum annorum 1772, 1773, 1774. et 1775. Viennæ et alibi locorum factarum / jussu Augustorum Dirigente *Maximiliano Hell*, astronomo cæsaro-regio universit. calculis definitæ a R. R. D. D. *Ignatio Lib. Barone de Rain* et *Francisco Güsman* astronomis universitatis. — Viennæ: Typis et sumptibus Joannis Thomæ de Trattner, Cæs. Reg. Maj. Aulæ Typhographi et Bibliopolæ, MDCCLXXV [1775]. — 266, [4], 53 p.

The following three items were publications by students:

8. **Exercitatio academica qua tribus satisfit de electricitate problematibus, unaque electrici fluidi actiones ad communes naturæ leges revocantur** / Edita communi nomine a D. D. *Antonio nobili de Graenzenstein*, *Philippo Krauss* et *Theodoro Zachariasiewicz*. — Leopoli : Typis Antonii Piller, Anno Salutis M.DCC.LXXIX [1779]. — 76 p.

**Theodor ZACHARIASIEWICZ** (Ukrainian: Теодор Захаріясевич; *1759 – †1808) was a priest and ecclesiastic figure of Halychyna. He served as a Vice-Rector of the Greek-Catholic General Seminarium in Lviv and a Professor of the Ecclesiastic History at Studium Ruthenum of the University of Lviv [Voznjak 1913; Kubijovyč 1955–57: II/2, p. 760].

9. **Exercitatio academica de sulphure et aqua Lubinensi. Quæ lustrarunt, et examini chemico subiecerunt** / *Bernardus de Ternowa Dwernicki*, *Valentinus Rutkowski* et *Thomas Twardochlebowicz* Physici leopolienses Assistente Francisco Güssmann Caes. Reg. Physices Professore. — [Leopoli, 1782]. — [17] f.
Bound with the following item.

10. **Tentamen Physicum de viribus in exiguis potissimum distantiis agentibus quod subirunt** / Perillustres, magnifici ac eruditi domini *Bernardus de Ternowa Dwernicki*, *Valentinus Rutkowski*, *Thomas Twardochlebowicz* Physici leopolienses Assistente *Francisco Güssmann* Caes. Reg. Physices Professore.— Leopoli: Typis Viduæ Josephæ Piller, Mense Julio Die […] Anno M.DCC.LXXXII [1782].— [7] f.
Bound with the previous item.

Both **Bernard DWERNICKI** and **Thomas TWARDOCHLEBOWICZ** (*? – †1806) became medical doctors [Kośmiński 1883: 100, 523] but biographical data are rather scarce.

11. **Beyträge zur Bestimmung des Alters unserer Erde, und ihrer Bewohner der Menschen** / von *Franz Güssmann*. — Wien : gedruckt und verlegt bey Joseph Gerold, 1782. — Erster Theil: Historische Beyträge zur Bestimmung der Epoche des menschlichen Geschlechtes. — 268 S.; [1783]. — [Zweyter Theil]. — 472 S.

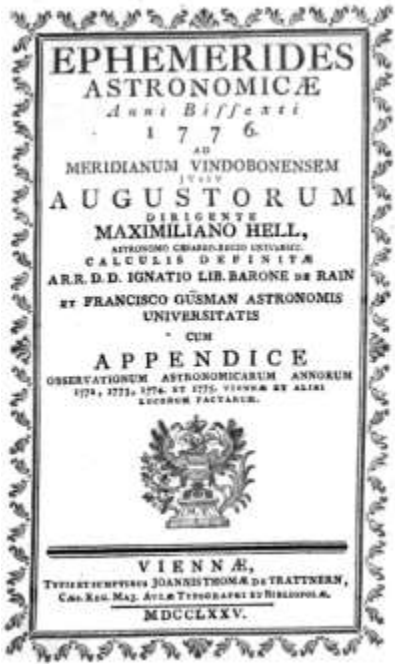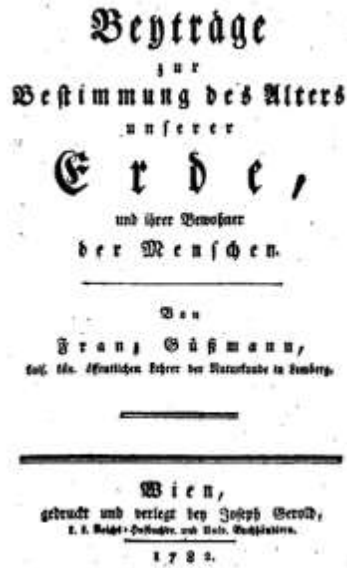

Figure 3: Title pages of works by Franz Güssmann. [Image source: GoogleBooks, MDZ]

**Franz GÜSSMANN** (*30.IX.1741, Wolkersdorf in Lower Austria – †28.I.1806, Seitenstetten in Lower Austria) was a physicist, member of the Jesuit Order. He taught physics at the University of Lviv in the 1770s–80s. In 1787 Güssmann became a Professor of Experimental Physics at the University of Vienna [WURZBACH 6, 20–21]. Item 7 above (*Ephemerides astronomicæ …*) is attributed to his Lviv period by Finkel and Starzyński [1894: I 40], as is the following work from 1785:

12. **Lithophylacium Mitisianum: dissertatione praeuia, et obseruationibus perpetuis physico mineralogicis explicatum** / a *Francisco Güssmann*. — Viennae : typis Iosephi Nobilis de Kurzbeck, 1785. — [6], 177, [1], 282, [2], 632 p., [1] h. pleg.

In this book, Franz Güssmann in particular expressed – for the first time ever – an idea that meteors fall from the sky [Marvin 1996].

Another work by Franz Güssmann from 1785 is the following title not being of a scientific nature:

13. **Tryphon, und Justin, oder vom Judentum**: Mit einer Vorrede an Herrn Moses Mendelssohn / von *Franz Güssmann*. — Wien : bey Stahel, 1785. — XIV, 227 S.

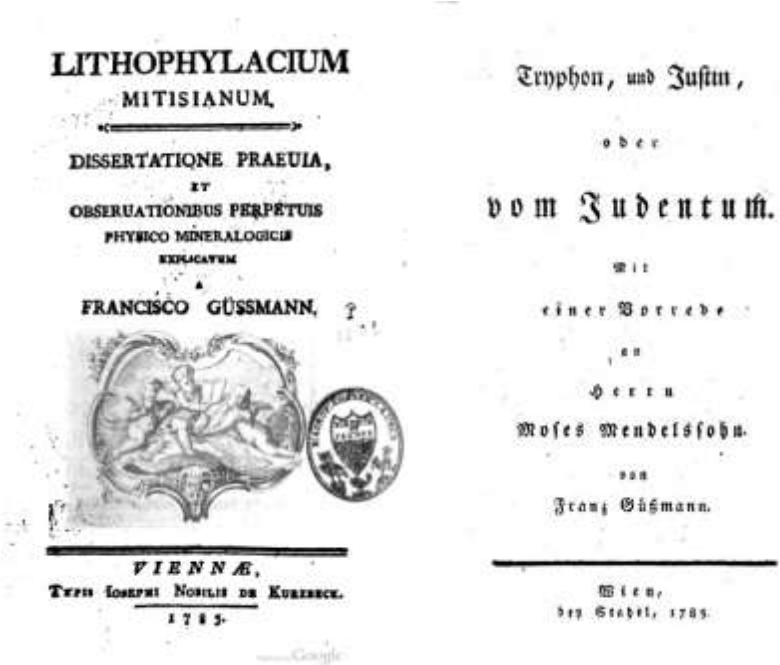

Figure 4: Title pages of books by Franz Güssmann from 1785. [Image source: GoogleBooks, MDZ]

On 21 October 1784 Emperor Joseph II issued a diploma about the foundation of a secular university in Lviv. This institution is known therefore as the Josephian University.

A newly created Chair of Physics was held by Ignác Martinovics.

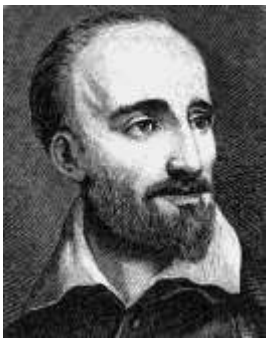 **Ignác Joseph MARTINOVICS** (*20 or 22.VII.1755 in Pest – †20.V.1795 in Buda) was a physicist, chemist, philosopher, and political adventurer. For some time he was a member of the Order of Franciscans having his religious name Dominicus. Ignác Martinovics studied theology in Buda graduating in 1779, the same year he received doctorate in theology and philosophy. He was a Professor of Physics and Mechanics at the University of Lviv in 1783–91. Prof. Martinovics served as a Dean of the Philosophical Faculty in 1785/86. In 1791 he obtained a laboratory in Vienna, then worked for some time in Szászvár. Ignác Martinovics initiated the Jacobin club in Pest; in 1795 he was arrested and beheaded in Buda [Sekulić 1978; WURZBACH 17, 50–56; Martinovics's file].

The first work by Martinovics, with possible attribution to the University of Lviv, could be

A. * **Dissertatio de harmonia naturali inter bonitatem divinam et mala creata, ad celeberrimam Hollandiae academiam Leidensem transmissa et nunc primum elucubrata** / *Ignatius Dominicus Martinowics*. — Leopoli, 1783. — [8], 49, [4] p.

This dissertation title, however, does not have any reference to the University of Lviv, as do most of his later works, so it probably should not be included in the bibliography and is thus given as supplementary information (hereafter this additional list is numbered A, B, C, etc.).

Since the following year and till 1792, numerous works by Martinovics are known.

## 1784

14. **Dissertatio de micrometro ope cuius vnus digitus geometricus diuiditur in 2,985,984 puncta quinti ordinis** / ab *Ignatio Martinovics*. — Leopoli : Typis Vidvae Iosephae Piller, M.DCC.LXXXIV [1784].— [11] f., 2 leaves of illustrations.

Another three works by students follow:

15. **Tentamen pvblicvm ex mathesi adplicata**, qvod in Caes. Regia Vniversitate Leopoliensi, die [31] Martii Anno M. DCC. LXXXIV svbiverunt, R. ac Ervd. D. Andreas Angełłowicz Alvm. Sem. Gener. Rit. Graeco-cath., Rel. ac Ervd. Fernandvs Syderowicz Min. Conv., Ervd. D. Antonivs Balinski, Ervd. D. Pavlvs Berezowski physicae in Annvm I. Avditores. — Leopoli : Typis Vidvae Iosephae Piller, [1784].— [6] f.

16. **Tentamen ex physica**, qvod die […] mensis Ivlii M.DCC.LXXXIV. svbivervnt Rev. ac Ervd. D. Andreas Angełłowicz, Sem. Gen. Rit. Graec. Cath. Alvmnvs. Ervd. D. Antonivs Balinski, Ervd. D. Pavlvs Berezowski. — Leopoli : Typis vidvae Iosephae Piller, Caes. Reg. Gvbern. Typograph, [1784].— [7] p.

17. **Tentamen ex physica**, quod die […] mensis Iulii M.DCC.LXXXIV svbivervnt Rev. et Ervd. D. Andreas Angełłowicz, Rev. et Ervd. D. Nicolaus Ivrkiewicz, Rev. et Ervd. D. Ioannes Sklepkiewicz, Sem. Gen. Rit. Graec. Cath. Alumni. Rev. et Ervd. D. Fernandvs[ ]Syderowicz, Min. Conv. — Leopoli : Typis Vidvae Iosephae Piller, [1784]. — [8] f.

**Andrij (Andreas) ANGEŁŁOWICZ** (Ukrainian: Андрій [де] Ангел[л]ович; *30.IX.1766 near Lviv – †1820) was a lawyer, theoretician of the ecclesiastical law. He taught law at the University of Lviv since 1789 both at Studium Ruthenum and Studium Latinum. In 1797 he became an extraordinary Professor. In 1800 Andrij Angełłowicz became a counselor of the nobility court in Tarnów [Kozytskyj & Pidkova 2007: 61].

## 1785

18. *Ignatii Iosephi Martinovics*, **Dissertatio physica de altitvdine atmosphaerae ex observationibvs astronomicis determinata, et anno MDCCLXXXV edita**. — Leopoli : Typis Pillerianis, [1785]. — [10] f., 1 page of illustrations.

19. **Tentamen publicum ex historia universali**, quod auctoritate et consensu spectabilis ac clarissimi domini *Ign. Jos. de Martinovics*… decani in… Universitate Leopolitana, secundum praelectiones Cl. D. *Zehnmark*… Subibunt domini auditores in annum primum & quidem I^{mo} D. Angełłowicz Joannes, D. Gorszinski Josephus, D. Kosinski Josephus, D. de Lasczewski Laurentius, 2^{do} D. Oslawski Valentinus, D. Poleiowski Mathias. D. Prominski Simeon, D. Stuorzinski Nicolaus. — [Leopoli] : Typis Pillerianis, M.DCC.LXXXV [1785]. — [4] f.

20. **Tentamen publicum ex historia universali**, quod auctoritate et consensu spectabilis ac clarissimi domini *Ign. Jos. de Martinovics*… decani… secundum praelectiones Cl. D. *Zehnmark*… Subibunt domini auditores in annum secundum R. ac Erud. D. Angełłowicz Simeon, R. ac Erud. D. Drosdowski Nicolaus, R. ac Erud. D. Kostecki Joannes Sem. gen. Rit. Graec. Cath. Alumni, Erud. D. de Levinski Joannes, R. ac. Erud. D. Lewinski Nicola[u]s quoque Sem. ejusdem Alumnus. — [Leopoli] : Typis Pillerianis, M.DCC.LXXXV [1785]. — [4] f.

A publication from a medicine graduate:

21. *Francisci Schravd Hvngari Pestani...* **Opvscvla rem physicam, et chemicam attinentia**: [Diss. I. De natura aerum inflammabilium et vitalium. II. De lucis et materiae electricae similitudine. III. De origine caloris in tritis corporibus. IV. De meteoro quodam singulari, et nonnullis consectariis theoriam aurorae borealis attinentibus]. — Leopoli : Typis Pillerianis, 1785. — [56] p.

    ESTREICHER 27, 261; SZÉCHÉNYI; WURZBACH 31, 274.

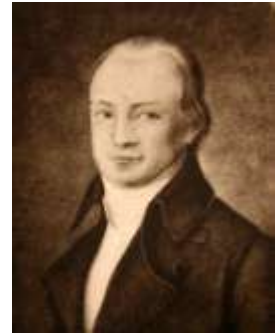

**Ferenc (Franz) SCHRAUD** (*14.V.1761 in Pest – †18.III.1806 in Kismarton or Vasvár) was a physician. He received his Doctor of Arts degree in Pest (1780), completed his medicine studies in Lviv and received the doctoral degree in Vienna. In 1790, Ferenc Schraud became the chief medical officer of Csongrád and Csanád counties. In 1792 he was called to head the Department of Theoretical Medicine of Surgeons. For his services during the outbreak of pest epidemic in Szerémség in 1794 he was ennobled, and was given the title of Royal Counselor for his services in the defense of pest in Bukovyna. He served as a Dean at the University of Buda in 1799/1800–1801/02. In 1802, Ferenc Schraud was appointed a National Head Physician (Protomedicus Hungariae) [WURZBACH 31, 272–274; Schraud n.d.].

## 1786

22. *Philosophische Fakultät. Die Professoren Vrecha, Rain, Martinovics, Zehnmark, Hann, Uhlich, Hiltenbrand, Umlauf, der Bibliothekar Brettschneider, und der Bibliothekkustos Kuralt,* **Sechster Brief** // Briefe über den itzigen Zustand von Galizien. Ein Beitrag zur Staatistik und Menschenkenntnis. — Erster Theil. — Leipzig: G. Ph. Wucherer, 1786. — S. 37-55.

## 1787

23. *Ign. Ios. Martinovics, Erudit. Scient. Societ. Electoralis Bavaricae. Hessen Hombovrg, Reg. Societ. Svecicae Soc. Membri, &c. In Reg. Scient. Vniversit. Leopolit. Physicae Experimentalis et Mechanicae Professoris Pvblici Ordinarii,* **Praelectiones physicae experimentalis**. — Tomvs Primvs. — Leopoli : Typis Thomae Piller, [1787]. — 349, [2] p., 2 leaves of illustrations; Tomvs Secvndvs. — 224 p., 4 leaves of illustrations.

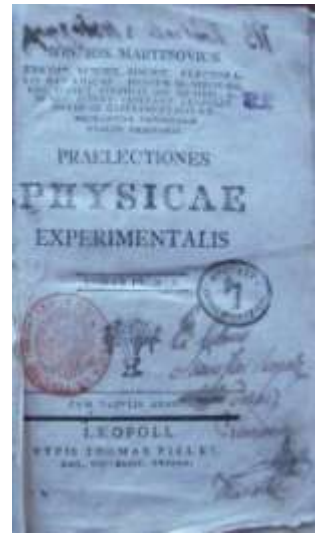

Figure 5: Title page of the first volume of *Praelectiones physicae experimentalis* by Ignac Martinovics
[Image source: LUL]

**1788**

24. **Memoires philosophiques, ou la nature devoilée** : Premiere partie / [*I. Martinovics*]. — à Londres, MDCCLXXXVIII [1788]. — 169, [3] p.

**1789**

25. \* **Physiologische Bemerkungen über den Menschen** / *Ignaz Josephs Martinovich*, k. k. Lehrers der Naturkunde und Mechanik auf der Hohen Schule zu Lemberg, der Chur-Fürst Bayerischen, Hessen-Homburgischen und königl. Schwedischen Gelehrten Gesellschaften Mitglieds. — Petropoli, 1789. — 71 p.

   PALLAS 12, 377; SZÉCHÉNYI provides a different spelling for the first word: **Phisiologische …**, as do some other sources [Fraknói 1921: 206; Balázs 2011]. According to [Balázs 2011], some copies appeared without the name of the author.

**1790**

26. *von Martinovich*, **Versuche über das Knallgold** // Beyträge zur Erweiterung der Chemie. — 1790. — B. 4, St. 2. — S. 149–155.

27. *Prof. von Martinovich*, **Salpeterartiges Bernsteinsalz** // Beyträge zur Erweiterung der Chemie. — 1790. — B. 4, St. 2. — S. 195–196.

28. *Martinovich*, **Fortgesetzte Versuche über das Knallgold** // Chemische Annalen…. — 1790. — Zweyter Theil, St. 8. — S. 98–103; St. 9. — S. 202–212.

29. *von Martinovich*, **Versuche über das Knallgold** // Taschen-Buch für Scheidekünstler und Apotheker. — auf das Jahr 1790: eifltes Jahr. — S. 104–114.

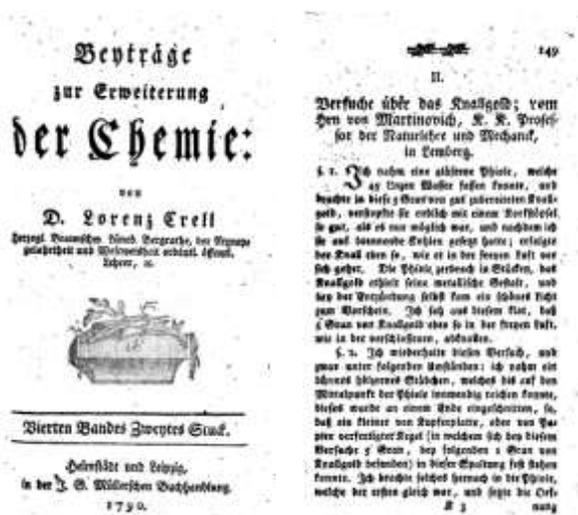

Figure 6: Ignac Matrinovics's paper from 1790 in *Beyträge zur Erweiterung der Chemie*

**1791**

30. *vom Hrn von Martinovich Prof. der Physik in Lemberg*, **Chemische Untersuchung des Galizischen Bergöhls** // Chemische Annalen…. — 1791. — Erster Band. — S. 32–29.

31. *vom Hrn Prof. I. I. von Martinovics*, **Chemische Abhandlung über die Grundstoffe der Laugensalze** // Chemische Annalen…. — 1791. — Zweiter Band, St. 9. — S. 196–206, 294–302.

An interesting example of early peer-review regarding a rejected submission by Martinovics from 1791 can be found in the journal of the St. Petersburg Academy of Sciences in Russia [Nova Acta 1795: 21]:

Le 28 Avril. Le Secrétaire a lu une lettre de M. Martinovich Profeſſeur à Leopol & datée du 10 Avril. M. Martinovitſch envoie pour être inſéré dans les Actes: Diſſertatio de nonnullis inſignioribus circuli proprietatibus, & déſire d'être agrée par l'Académie au nombre de ſes aſſociés étrangers. Comme cette Diſſertation contient la réfutation d'une prétendue quadrature de cercle, & que d'ailleurs on n'y trouve aucune propoſition nouvelle & remarquable du cercle, comme le titre paroit le promettre, la lettre de M. Martinovitſch reſta ſans réponſe, & ſon mémoire fut mis de rebut.

A contribution with a similar title was published in *Göttingische Anzeigen*, see item 37.

Figure 7: Ignac Matrinovics's paper from 1791 in *Chemische Annalen*

Figure 8: Ignac Matrinovics's paper from 1791 in *Taschen-Buch für Scheidekünstler*

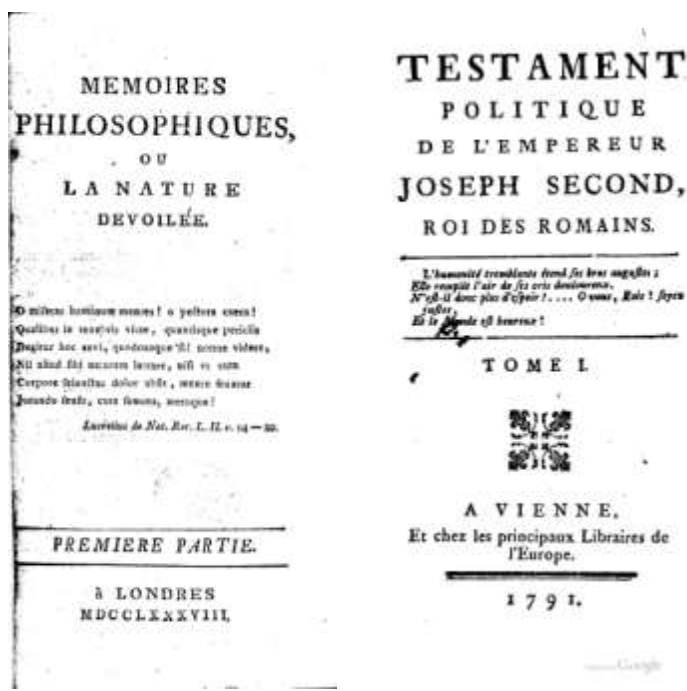

Figure 9: Title pages of the anonymous works written by Ignac Martinovics. [Image source: MEK, GoogleBooks]

## 1792

39. *Prof. I. I. von Martinowich*, **Ueber den Ursprung der im Wasser befindlichen Luft** // Beyträge zur Erweiterung der Chemie. — 1792. — B. 5, St. 3. — S. 267–274.

40. *Prof. von Martinovich*, **[Auszug aus einem Brief]** // Taschen-Buch für Scheidekünstler und Apotheker. — auf das Jahr 1792: Dreizehntes Jahr. — S. 186–189.

41. *Martinovich*, **Observations sur une espèce de Pétrole qui contient du Sel sédatif** // Observations périodiques sur la Physique, sur l'Histoire Naturelle et sur les Arts. — 1792. — Tome XL, Part. I, avril. — P. 315–316.

42. *Martinovich*, **Observations sur une espece de pétrole qui contient du sel sédatif** // Journal encyclopédique ou universel. — 1792. — Tome IV, dix juin, Nº XVI. — P. 543–545.

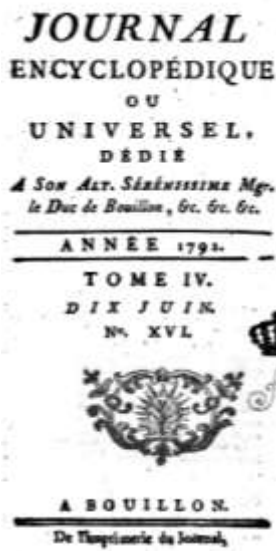

Figure 10: Ignac Martinovics's paper from 1791 in *Journal encyclopédique ou universel*

After Martinovics left the University, the Chair was subsequently held by Anton Hiltenbrandt in 1792–1794 and Ivan (Jan, Johannes) Zemantsek in 1794–1805.

**Anton HILTENBRAND** (*1721, Vienna – †25.VIII.1798, Vienna) – physicist and geographer, He was a Professor of philosophy, history and physical geography in Vienna. In 1784–92 he was a Professor of Natural History, in 1792–94 a Professor of Physics at the University of Lviv. In 1794 Prof. Hiltenbrand was ennobled for his merits. Anton Hiltenbrand did not publish any work while being a Professor of Physics at the University of Lviv (but see item 22 above). The attribution of the $3^{rd}$ edition of his *Österreichischer Weinbaukatechismus oder kurzer Unterricht vom Weinbaue in Österreich* to 1793 by Finkel and Starzyński [1894: I 68] seems to be a typo as other sources give 1796 [WURZBACH 9, 133].

**Ivan (Jan, Johannes) Nepomuk ZEMANTSEK** (Zemánček) (*1759, Transcarpathia? – †19.IV.1825, Vienna) was a naturalist, physicist, mathematician, doctor of philosophy. In 1786–87 he was an adjunct of higher mathematics in Budapest. In 1787–1805 Ivan Zemantsek taught physics at the University of Lviv (since 1794 as a professor), both in Latin and Ukrainian languages (at Studium Ruthenum). In 1795/96 he was Dean of the Philosophical Faculty and in 1803/04 served as Rector of the University of Lviv. Upon the University closure, he was a Professor of Physics at the Jagiellonian University in Cracow and in 1806/07 served as Dean of the Philosophical Faculty there. In 1811–12 Ivan Zemantsek was a Professor of Physics at the Lyceum in Linz and in 1813–22 was a Professor of Physics and Mechanics at the University of Vienna. No published works by Prof. Zemantsek are known [Voznjak 1913; Encyclopedia 2011: 533].

# 4. Franzian University

By the end of 1805, the University of Lviv was moved to Cracow according to an order by Kaiser Franz II from 09 August 1805. The institution in Lviv was converted into Lyceum, which was similar to the University from the organizational point of view. The Chair of Physics remained a structural unit within philosophical studies. The Lyceum functioned in 1805–1817, and on 17 August 1817 the University was re-established in Lviv, according to Emperor Franz I decision.

The Chair of Physics was first held by Anton Gloinser, who did not publish. Being rather a good educator than a scientist, he paid special attention to teaching. The Lviv University Library holds an almost 400-page manuscript of Gloisner's lectures from the academic year 1813/14 entitled *Physica…* (see Fig. 11), which are based mostly on the handbook *Elementa physicae mathematico-experimentalis* (Viennae: Geissinger, 1812) by Remigius Döttler, Professor of Physics in Vienna [Finkel & Starzyński 1894: I 228].

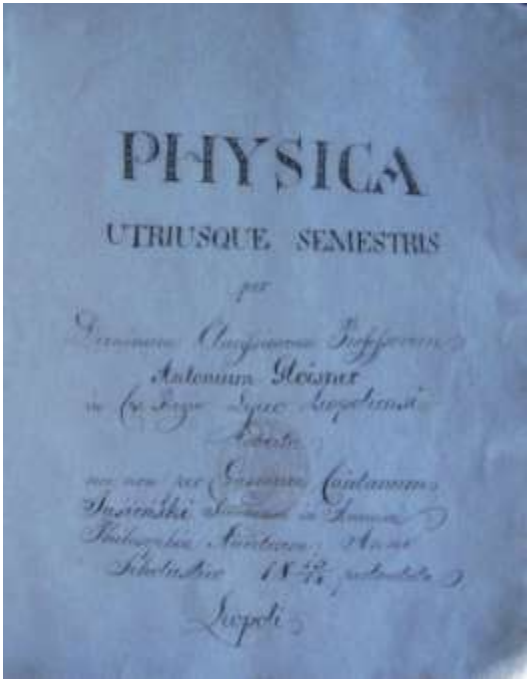

Figure 11: *Physica…*, a manuscript of lectures by Anton Gloisner. [Image source: LUL]

**Anton GLOISNER** (*1782 in Halychyna – †II.1855 in Lviv) was a physicist, botanist, and mathematician. He graduated from the University of Lviv in 1805. Anton Gloisner headed the Chair of Physics in 1807–23, in 1809–13 an ordinary Professor of Physics and extraordinary Professor of History and Technology, in 1809–16 was a Professor of Natural History. In 1818, 1819/20 he was a Dean of the Philosophical Faculty at the University of Lviv. Anton Gloisner stopped teaching in 1823 because of deafness and was for some time a counselor of the nobility court in Stanislaviv and later in Lviv. [Gloisner's file; Finkel & Starzyński 1894: I 228; Svojtka 2010].

Gloisner's successor, August Kunzek, was not only a scientist with a wide outlook, but also a good educator and popularizer of science.

**August KUNZEK, Edler von Lichton** (*28.I.1795, Königsberg in Schlesien (Klimkovice) – †31.III.1865, ?) was a physicist. He studied philosophy at the University of Olomouc in (1815–17) and later law, mathematics, and physics at the University of Vienna. In 1822–24 he was an Adjunct of Physics and Mathematics and since 1847 a Professor of Physics at the University of Vienna. In 1824–48, August Kunzek was a Professor of Physics and also since 1827 an extraordinary Professor of Popular Mechanics at the University of Lviv. In

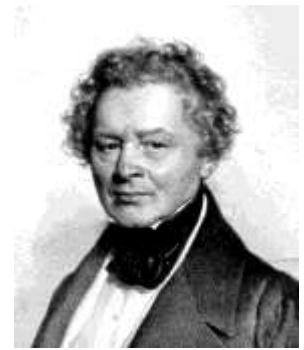

1827/28 and 1840/41 he served as Dean of the Philosophical Faculty and in 1832/33 was a Rector of the University of Lviv. Professor Kunzek was a Corresponding Member of the Academy of Sciences in Vienna (1848). In 1848 he was awarded the Order of the Zähringer Lion. In 1862 August Kunzek was awarded the nobility title Edler von Lichton [OeBL 4, 357; Wurzbach 13, 390–392; Kunzek's file]. Portrait courtesy of the Austrian National Library.

**1836**

43. **Die Lehre vom Lichte nach dem neuesten Zustande der Wissenschaft zunächst für das Bedürfniss gebildeter Stände** / dargestellt von *August Kunzek*. — Lemberg, Stanislawow und Tarnow : Verlag von Joh. Millikowski, 1836. — 449, [7] S., 5 gef. Bl.

**1842**

44. **Leichtfassliche Vorlesungen über Astronomie**: für jene, denen es an matematischen Vorkenntnissen fehlt / von *August Kunzek*. — Wien : Verlag von Ignaz Klang, 1842. — VIII, 218 S.

**1847**

45. **Leichtfassliche Darstellung der Meteorologie** / von *August Kunzek*. — Wien : Braumüller und Seidel, 1847.— [4], 276 S.

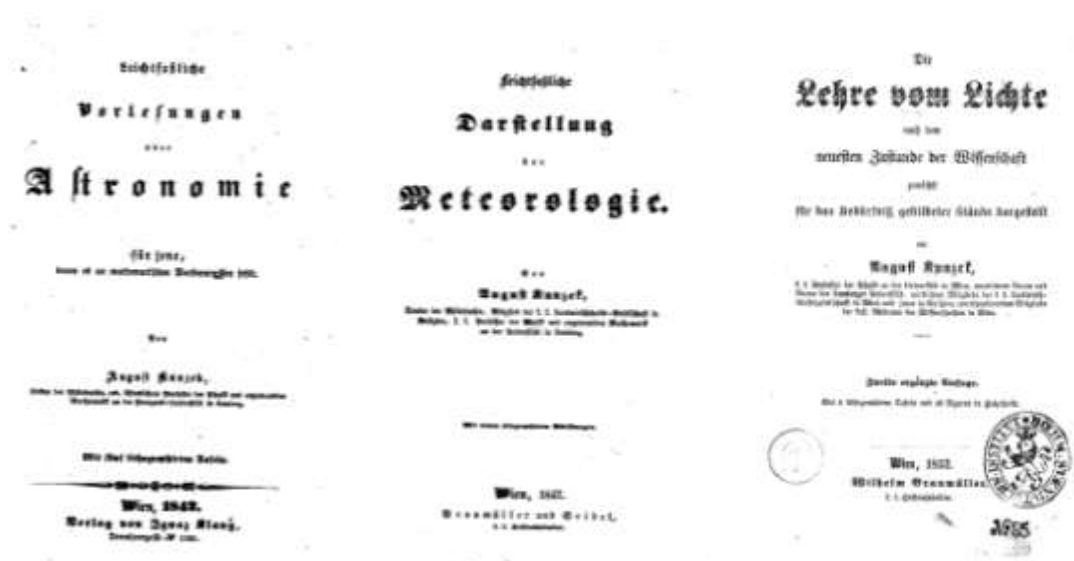

Figure 12: Title pages of books by August Kunzek. [Image source: GoogleBooks]

In 1849, the Polish translation of Kunzek's work entitled *News from the field of physics, chemistry and mechanics...* was published. The translation was made by Wojciech Urbański, who became a docent of mathematical physics a year later. This free translation bears on the title page a note about August Kunzek referring him to as a former professor of Physics at the University of Lviv. While it seems quite possible that the work had been written yet during his Lviv period, it is given here in the supplementary list only:

B. **Wiadomości z fizyki, chemii i mechaniki** dla użytku gospodarzy wiejskich / przez Augusta Kunzeka ; wolny przekład z niemieckiego [przez Dr. Wojciecha Urbańskiego]. — we Lwowie : nakładem Towarzystwa gospodarskiego ; z drukarni Zakładu narodowego Ossolińskich, 1849. — 90, [2] s.

After August Kunzek, the Chair of Physics was consecutively held by Aleksander Zawadzki, Viktor Pierre, Wojciech (Adalbert) Urbański, and Alois Handl.

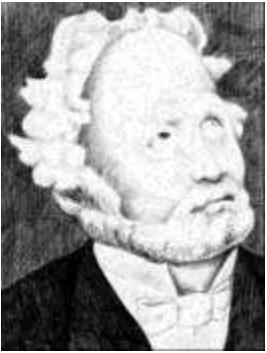

**Aleksander ZAWADZKI** (*06.V.1798, Bielsko, now a part of Bielsko-Biała in Poland – †06.V.1868, Brno in Moravia) was a naturalist, mostly working in zoology and botany. He studied at the University of Olomouc in 1815–17 and at the University of Lviv in 1818–26. He obtained his Doctor of Philosophy degree in 1829. In 1828–35 he was a teacher at the Seminarium in Lviv, in 1835–37 taught botany and physics for medical students at the University of Lviv. In 1837 he became Professor of Physics and Mathematics at the Philosophical Institute in Przemyśl. In 1849–53, Zawadzki held the Chair of Physics, in 1850/51 was a Dean of the Philosophical Faculty at the University of Lviv (fired for supporting radical students and professors during the revolution in 1848). In 1853 he became a Director of the Technical School in Brno. Aleksander Zawadzki was a member of numerous scientific societies. Endemic plant *Chrysantemum zawadzkii* Herb. (modern *Tanacetum zavadzkii*) and several beetles (*Carabus zawadzkii* Frivaldszky and *Dryomyza zawadzkii* Schummel) are named after him. [Syniawa 2000; Encyclopedia 2011: 506; Zawadzki's file].

In fact, Aleksander Zawadzki did not publish in physics. Two of the works, which appeared when he held the Chair of Physics, are on paleontology.

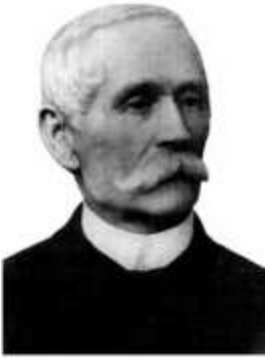

**Wojciech (Adalbert) URBAŃSKI** (*28.III.1820 in Khodoriv, now Lviv Oblast, Ukraine – †25.VI.1903, Lviv?) was a physicist. He studied in Berezhany, Stanislaviv, Ternopil, Lviv, and Vienna, where he obtained his Doctor of Philosophy degree in 1847. Wojciech Urbański was then a teaching philosophy, mathematics, and physics at the Gymnasium in Przemyśl. Since 1849 he worked as a "scriptor" and since 1852 as a Curator of the Lviv University Library. In 1850 Wojciech Urbański habilitated as a Docent of Mathematical Physics and started lecturing at the University of Lviv, since 1857 as an acting Professor of Physics. In 1859–1892, he was a Director of the Lviv University Library [WURZBACH 49, 129–133; Dąbrowski 2007; Urbański's file].

## 1850

46. *Dr. Adalbert Urbański* **Abhandlung über ein Problem aus der Elektrostatik** // Erstes Programm des k. k. akademischen Ober-Gymnasiums in Lemberg am Schluss des Schuljahres 1850. — Lemberg, [1850]. — S. 1–33.

47. **Pogląd ogólowy na dzisiejsze stanowisko nauki o tak zwanych imponderabiliach** / przez *Dr. A. E. Urbańskiego* // Pamiętnik literacki…. — 1850. — Rok I, Nr. 4. (20 kwietnia). — S. 73–79.

48. **Proces oddechania, skreślony ze stanowiska umiejętności dzisiejszej** / przez *Dra A. E. Urbańskiego* // Pamiętnik literacki…. — 1850. — Rok I, Nr. 8. (24 maja). — S. 169–180.

49. *Dr. Alexander Zawadzki* **Ueber die Wichtigkeit der Paläontologie, oder Versteinerungskunde** // Jahresbericht des k. k. Obergymnasiums bei den Dominikanern in Lemberg für das Schuljahr 1849/50. — Lemberg : Aus der k. k. galizischen Provinzial-Staats-Druckerei, [1850]. — S. 3–17.

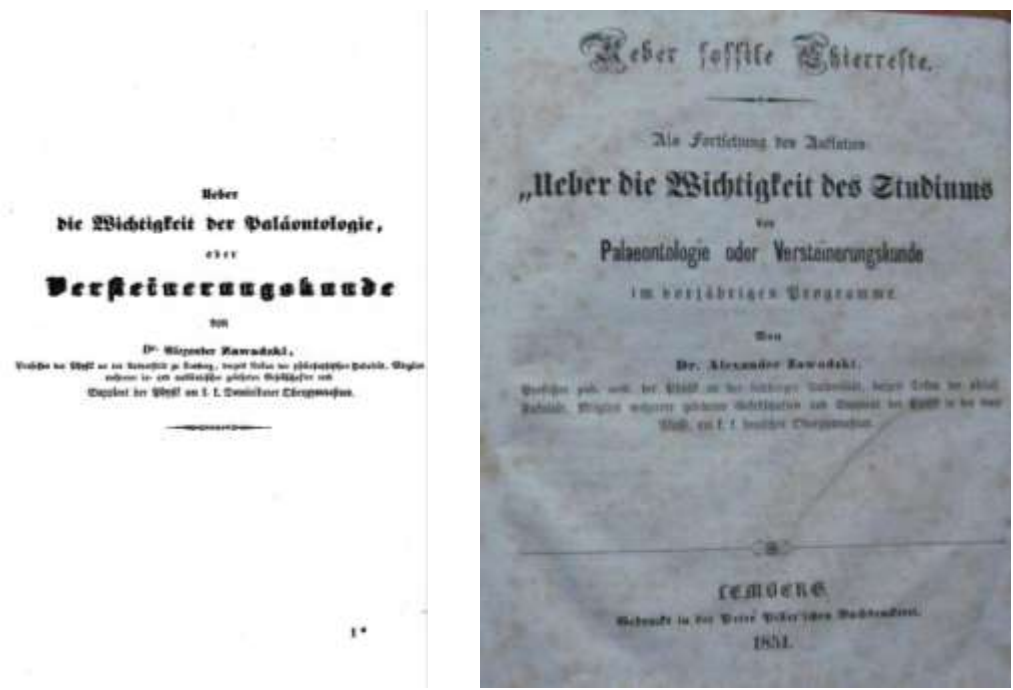

Figure 13: Title pages of works by Aleksander Zawadzkii. [Image sources: FBC, LUL]

## 1851

50. *Dr. Alexander Zawadzki*, **Ueber fossile Thierreste** : als Fortsetzung des Aufsatzes: "Ueber die Wichtigkeit des Studiums der Palaeontologie oder Versteinerungskunde im vorjährigen Programme // Jahres-Bericht des kaiserl. königl. deutschen Ober-Gymnasiums und der damit verbundenen polnischen vier Parallel-Classen bei den Dominikanern in Lemberg für das Schuljahr 1850/51. — Lemberg : Gedruckt in der Peter Piller'schen Buchdruckerei, 1851. — S. 1–16.

51. **Fizyka na trzecią klasę w gimnazyach niższych** / przez *Wojciecha Urbańskiego*. — Lwów : w drukarni Zakładu Narodowego Ossolińskich, 1851. — [2], 78 s.

52. **Gieometryja wyłożona sposobem uzmysławiającym do użytku młodzieży w gimnazyjach niższych** / przez *Dra. A. E. Urbańskiego*. — Lwów : Nakładem P. Stockmana, 1851. — [2], 247 s.

53. *Dr. Adalbert Urbański,* **Cantate über die Ankunft Seiner Majestät des Kaisers Franz Josef in Lemberg**. — Lemberg : Druck aus Peter Piller's Buchdrukerei, [1851]. — [4] S.

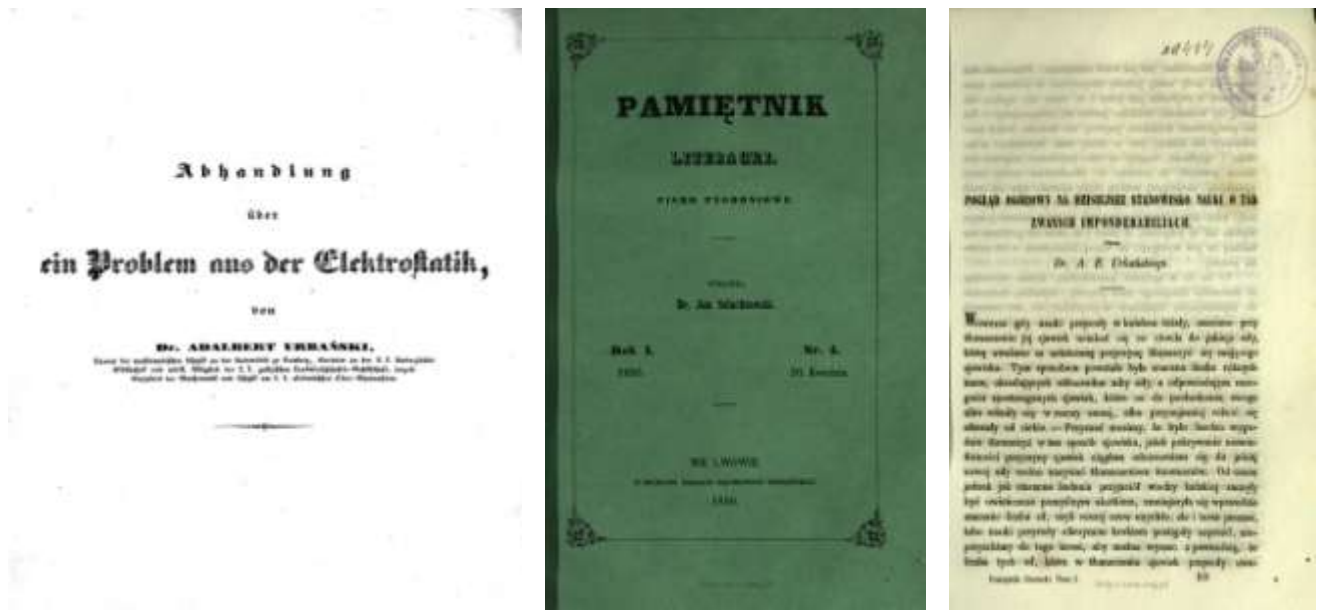

Figure 14: Works by Wojciech Urbański from 1850. [Image sources: FBC]

**Viktor PIERRE** (\*19.XII.1819 in Vienna – †29.X.1886 in Vienna) was a physicist. He was studying physics and medicine at the University of Vienna (Doctor of Medicine in 1844; Doctor of Philosophy in 1846). In 1844 he was an Adjunct of Pure Mathematics and Physics, in 1848 Supplent of Physics and Adjunct of Mathematics and Physics at the University of Vienna. In 1851 Viktor Pierre become a Professor of Physics at the Technical Academy in Lviv. In 1853–57 he was a Professor of Physics at the University of Lviv. In 1857 he moved to the University of Prague and since 1866 was a Professor of Physics at the Polytechnic Institute of Vienna (since 1872 Technische Hochschule) and served as a Rector in 1873/74 [OEBL 8, 66–67; Pierre's file].

### 1854

54. *Dr. Victor Pierre,* **Beitrag zur Theorie der Gaugain'schen Tangentenbussole** // Sitzungsberichte der Kaiserlichen Akademie der Wissenschaften. Mathematisch-naturwissenschaftlichen Classe.— 1854.— B. 13, Hft II.— S. 527-531.

### 1855

55. *Dr. Victor Pierre,* **Beitrag zur Theorie der Gaugain'schen Tangentenbussole** // Annalen der Physik und Chemie.— 1855.— Vierte Reihe Vierter Band; Band XCIV; der ganzen Folge hundert und siebzigter.— S. 165-169.

56. **O cholerze i sposobie powstrzymania onéjże** / [napisał *Dr. Wojciech Urbański*]. — Lwów : Drukiem Piotra Pillera i Syna, 1855.— 15 s.

### 1856

57. *J. Pless, Victor Pierre,* **Beiträge zur Kenntnis des Ozons und des Ozongehaltes der atmosphärischen Luft** // Sitzungsberichte der kaiserl. Akademie d. Wissenschaften in Wien math.-naturw. Classe. — 1856. — Band 22. — S. 211–235.

58. *Dr. Wojciech Urbański*, **Proces oddychania ze stanowiska chemiczno-fizjologicznego** // Przyroda i Przemysł: tygodnik… (Poznań). — 1856. — Rok 1, № 35. — S. 277–279; № 36. — S. 285–288; № 37. — S. 293–295.

Figure 15: Papers by Viktor Pierre. [Image souces: Gallica, BDHL]

## 1857

59. *Dr. Wojciech Urbański*, **O warunkach rozwijania się roślin** // Przyroda i Przemysł: tygodnik… (Poznań). — 1857. — Rok 2, № 1. — S. 1–4; № 2. — S. 9–11; № 3. — S. 17–20; № 4. — S. 25–28; № 5. — S. 33–35; № 6. — S. 41–43.

60. *Dr. Wojciech Urbański*, **O kometach** // Przyroda i Przemysł: tygodnik… (Poznań). — 1857. — Rok 2, № 18. — S. 137–139; № 19. — S. 145–147; № 20. — S. 153–155; № 21. — S. 161–163; № 22. — S. 169–172; № 28. — S. 217–219; № 39. — S. 305–309.

61. **Vorträge über höhere Physik** gehalten an der k. k. Lemberger Hochschule in den Jahren 1851 bis 1856 / vom Privatdocenten *Dr. Adalbert Urbański*. — 1. Abtheilung: Polarwirkende Naturagentien. — Lemberg : In der Buchdruckerei des Kornel Piller, 1857. — VIII, 142, [1] S.

Figure 16: Some works by Wojciech Urbański [Image sources: FBC, LUL, MDZ]

## 1858

62. **Magnetische Beobachtungen in Lemberg, ausgeführt im Monate October 1858** / von
*Dr. Adalbert Urbański*. — Lemberg : Kornel Piller, 1858. — 40 S.

63. **O kometach** / przez *Dra Wojciecha Urbańskiego*. — Poznań : Ludwik Merzbach, 1858. — [1], 65 s.

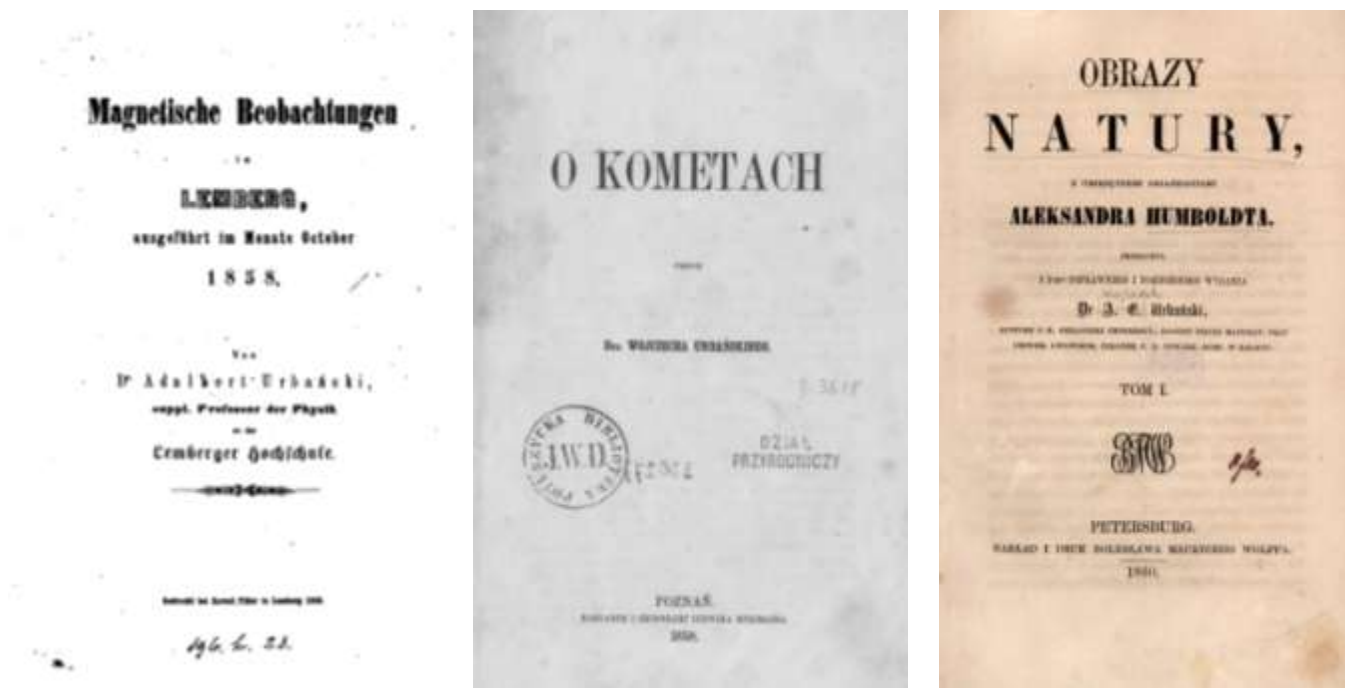

Figure 17: Title pages of the books by Wojciech Urbański. [Image sources: GoogleBooks, FBC, Polona]

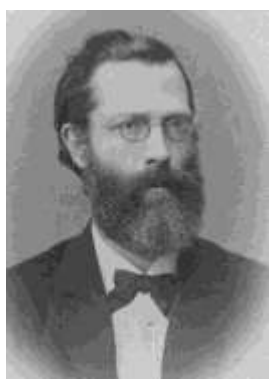

**Alois HANDL** (*22.VII.1837, Feldkirch in Vorarlberg, – †10.II.1915, Suceava in Bukovyna, now Romania) was a physicist, doctor of philosophy (1859). He studied in Vienna and Graz. In 1852–62 Alois Handl was a deputy Professor, and in 1862–72 ordinary Professor of Physics at the University of Lviv. In 1870/71 he was a Dean of the Philosophical Faculty. In 1872 Prof. Handl moved to Wiener-Neustadt, where he was a Professor at the Military Academy, and in 1876–1906 he worked as a Professor of Experimental Physics at the University of Chernivtsi, serving in 1894/95 as Rector. Alois Handl was a member of the German Leopoldina Academy in Halle. See for more details [Finkel & Starzyński 1894: I 324; Handl's files].

**1859**

64. *Alois Handl*, **Über die Krystallformen einiger chemischen Verbindungen** // Sitzungsberichte der kaiserl. Akademie d. Wissenschaften in Wien math.-naturw. Classe. — 1859. — B. 37. — S. 386–392, 1 Tafel.

**1860**

65. **Obrazy natury** : z umiejętnymi objaśnieniami Aleksandra Humboldta / przełożył z 3-go poprawnego i pomnożonego wydania *Dr. A. E. Urbański*. — Petersburg : Nakład i druk Bolesława Maurycego Wolfa, 1860. — Tom I. — [2], VI, 246 s.; Tom II. — [4], 279 s.

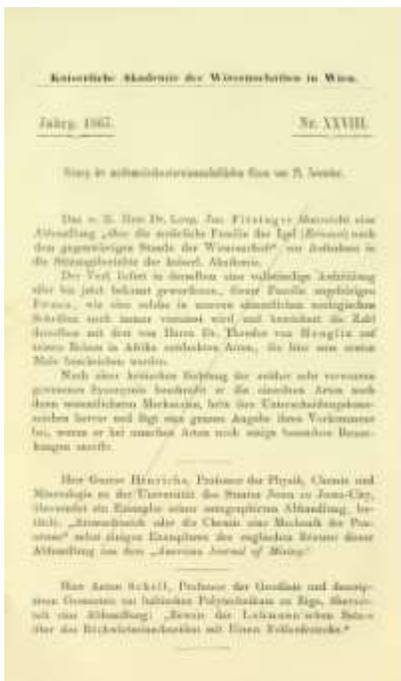
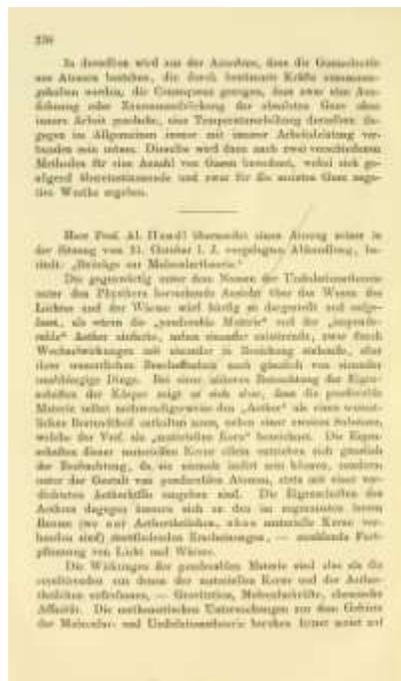

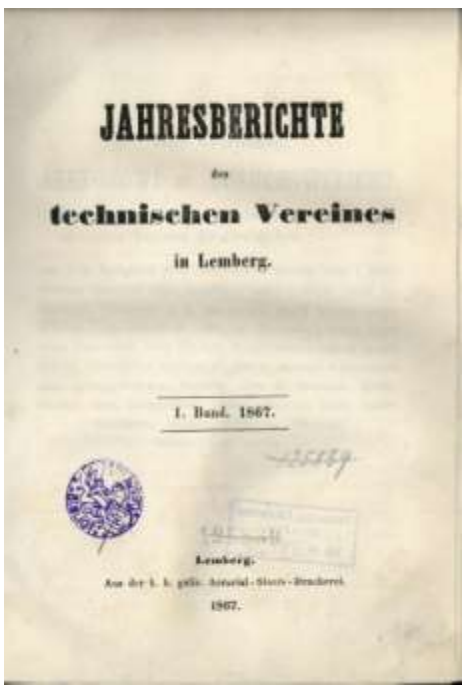
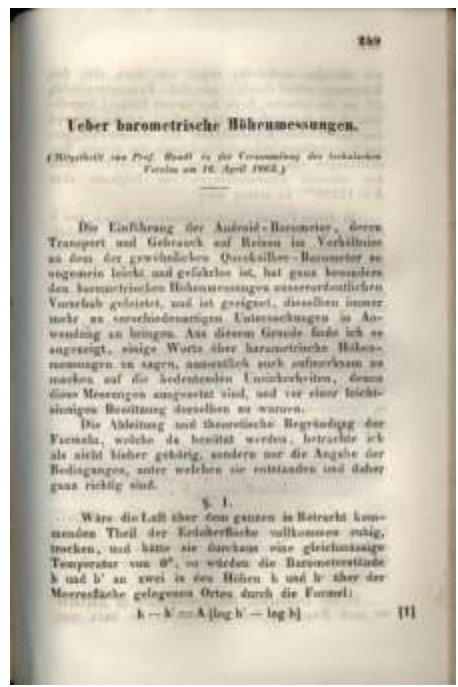

Figure 18: Papers by Alois Handl from 1867. [Image sources: BDHL, LSL]

## 1868

74. *Dr. A. Handl*, **Über eine neue Art der Beobachtung an Heberbarometern** // Sitzungsberichte der Kaiserlichen Akademie der Wissenschaften. Mathematisch-naturwissenschaftlichen Classe.— 1868. — B. 57, II. Abt. — S. 109–114.

75. *Prof. Al. Handl*, **Beobachtungen am Heberbarometer** // Anzeiger der Österreichischen Akademie der Wissenschaften, Mathematisch-Naturwissenschaftliche Klasse. — 1868. — V. Jahrgang, Nr. IV. — S. 33–34.

## 1869

76. *Dr. A. Handl*, **Theorie der Waagebarometer** // Sitzungsberichte der Kaiserlichen Akademie der Wissenschaften. Mathematisch-naturwissenschaftlichen Classe. — 1869. — B. 59, II. Abt. — S. 7–16.

77. *Alois Handl*, **Theorie der Waagebarometer** // Anzeiger der Österreichischen Akademie der Wissenschaften, Mathematisch-Naturwissenschaftliche Klasse. — 1869. — VI. Jahrgang, Nr. I. — S. 4.

78. *Dr. Alois Handl*, **Ueber eine neue Art der Beobachtung and Heberbarometern** // Repertorium für Experimental-Physik, für physikalische Technik, mathematische und astronomische Instrumentenkunde. — 1869. — B. 5. — S. 30–35.

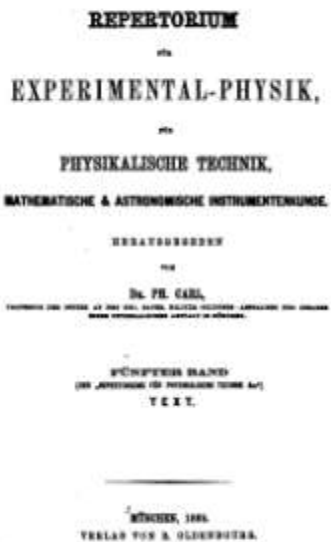 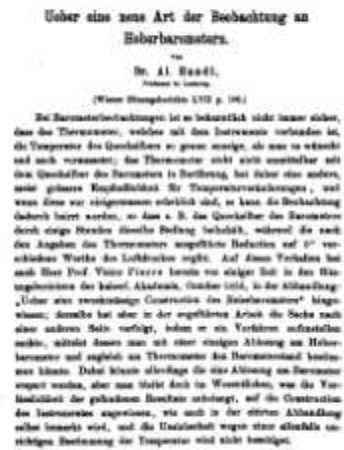

Figure 19: Alois Handl's paper from 1869 [Image source: GoogleBooks]

## 1870

79. *Dr. Alois Handl*, **Theorie der Wagebarometer** // Repertorium für Experimental-Physik, für physikalische Technik, mathematische und astronomische Instrumentenkunde. — 1870. — B. 6. — S. 104–112.

## 1871

80. *Dr. A. Handl*, **Notiz über die alteren meteorologischen Beobachtungen in Lemberg** // Sitzungsberichte der Kaiserlichen Akademie der Wissenschaften, mathematisch-naturwissenschaftlichen Classe. — 1871. — B. 64, Abtheilung 2. — S. 57–61.

81. * *Alois Handl*, **Uebersicht über die Temperaturs-Verhältnisse in Lemberg, im Zeitraume 1824-1870**. — Lemberg: Verlag des Technischen Vereins, [1871]. — [12] S.; with a two-page translation into Polish by S. P.: **Przegląd zmian temperatury we Lwowie od r. 1824 aż do r. 1870** / tłumaczył S. P. — Lwów : druk. E. Winiarza. — 2 s.

ESTREICHER II ed. 10, 56.

## 1872

82. *Dr. Alois Handl*, **Notiz über absolute Intensität und Absorption des Lichtes** // Sitzungsberichte der Kaiserlichen Akademie der Wissenschaften, mathematisch-naturwissenschaftlichen Classe.— 1872. — B. 65. — S. 129–132.

83. *Dr. Alois Handl*, **Über die Constitution der Flüssigkeiten (Beiträge zur Molekulartheorie. II.)** // Sitzungsberichte der Kaiserlichen Akademie der Wissenschaften, mathematisch-naturwissenschaftlichen Classe. — 1872. — B. 65. — S. 377–388.

84. *Dr. Al. Handl*, **Über den Zustand gesättigter und übersättigter Lösungen (Beiträge zur Molekulartheorie. III.)** // Sitzungsberichte der Kaiserlichen Akademie der Wissenschaften, mathematisch-naturwissenschaftlichen Classe. — 1872. — B. 66. — S. 136–142.

85. *Handl*, **Ueber den Zustand gesättigter und übersättigter Lösungen** // Repertorium für Experimental-Physik, für physikalische Technik, mathematische und astronomische Instrumentenkunde. — 1872. — B. 8. — S. 379.

86. *Prof. Handl*, **Ueber absolute Intensität und Absorption des Lichtes** // Anzeiger der Kaiserlichen Akademie der Wissenschaften, Mathematisch-Naturwissenschaftliche Klasse. — 1872. — IX. Jahrgang, Nr. VIII. — S. 50.

87. *Prof. Dr. Al. Handl*, **Über die Constitution der Flüssigkeiten** // Anzeiger der Kaiserlichen Akademie der Wissenschaften, Mathematisch-Naturwissenschaftliche Klasse. — 1872. — IX. Jahrgang, Nr. XIII. — S. 88.

88. *Prof. Al. Handl*, **Über den Zustand gesättigter und übersättigter Lösungen** // Anzeiger der Kaiserlichen Akademie der Wissenschaften, Mathematisch-Naturwissenschaftliche Klasse. — 1872. — IX. Jahrgang, Nr. XIX. — S. 125.

89. *Dr. Al. H.* **Ueber den Welt-Aether und die Naturkräfte** // Neue Freie Presse. Abendblatt. — Wien, Donnerstag, den 27. Juni 1872. — S. 4.

90. *Dr. Al. H.* **Ueber den Welt-Aether und seine Wirkungen** // Neue Freie Presse. Abendblatt. — Wien, Donnerstag, den 18. Juli 1872. — S. 4.

91. *Dr. Al. H.* **Ueber den Welt-Aether und die Materie** // Neue Freie Presse. Abendblatt. — Wien, Donnerstag, den 12. September 1872. — S. 4.

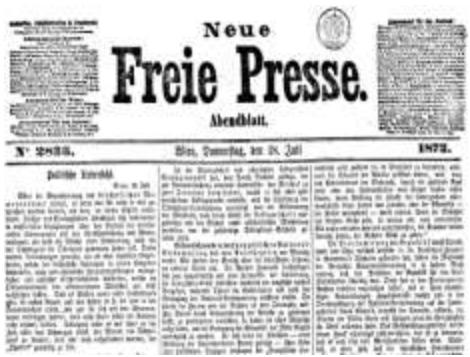

Figure 20: One of the last Handl's work from the Lviv period printed in *Neue Freie Presse*, a newspaper in Vienna. [Image source: ANNO]

The year of 1872 was the final year in the almost centenary history of the Chair of Physics at the University of Lviv. The demands of time led to a more specific specialization between theorists and experimentalists. In 1872, two new physicists habilitated as docents at the University. Next year, separate professorships in Theoretical and Experimental Physics were created and thus the Chair of Physics split into the Chair of Theoretical Physics (held by Oskar Fabian) and Chair of Experimental Physics (held by Tomasz Stanecki) [Finkel & Starzyński 1894: II130; Rovenchak 2009].

## 5. Summary

In the work, a detailed bibliography related to physics at the University of Lviv in 18th–19th centuries is presented. Brief biographical accounts of the authors are given to put their works in the context of the University history.

The bibliography list contains 91 items having been discovered so far; nine of them were not checked *de visu*. Two most prolific authors were Alois Handl (26 titles) and Ignac Martinovics (25 titles). Wojciech Urbański authored 15 publications, Franz Güssmann was an author of four items plus he is mentioned in one publication by his assistants. August Kunzek was an author of three books and Viktor Pierre authored three papers during their work at the University. Two items (on paleontology) were published by Aleksander Zawadzki. The authorship of four publications is not known. Andreas Angełłowicz was an author of three items with several other coauthors. Two items were authored jointly by Dwernicki, Rutkowski, and Twardochlebowicz and one item jointly by Graenzenstein, Krauss, and Zachariasiewicz. Finally, Kajetan Tęgoborski, Ludwik Hoszowski, and Ferenc Schraud authored one publication each.

This is the first study of such a sort related to natural sciences at the University of Lviv and – to my knowledge – in the history of this academic tradition in Central-Eastern Europe. The analyzed materials will facilitate research in the field of the history of physics and the bibliography of natural sciences.

Kunzek's file: State Archive of Lviv Oblast. Repository 26, Desc. 5, Case 1012.

Martinovics's file: State Archive of Lviv Oblast. Repository 26, Desc. 5, Case 1208.

Pierre's file: State Archive of Lviv Oblast. Repository 26, Desc. 5, Case 1580.

Urbański's file: State Archive of Lviv Oblast. Repository 26, Desc. 5, Case 1927.

Zawadzki's file: State Archive of Lviv Oblast. Repository 26, Desc. 5, Case 694.

**Bibliographies, encyclopedias, and library catalogs:**

**Digital libraries and repositories:**

Internet Archive, https://archive.org

MEK (Magyar Elektronikus Könyvtár): http://mek.oszk.hu

MDZ (Münchener Digitalisierungszentrum): http://www.digitale-sammlungen.de

Oberösterreichisches Landesmuseum: http://www.landesmuseum.at

Polona: http://polona.pl